\documentclass[11pt,a4paper]{article}
\usepackage{jheppub}
\usepackage{graphicx}
\usepackage{epsfig}
\usepackage{amsmath}
\usepackage{amssymb}
\usepackage{amsthm}
\usepackage{ulem}
\usepackage{color}
\usepackage[utf8]{inputenc}

\newcommand{\tr}{\text{Tr}}

\title{On BPS Equations of Generalized $SU(2)$ Yang-Mills-Higgs Model with Scalars-Dependent Coupling $\theta$-term}
\author{Mulyanto, Emir Syahreza Fadhilla, Ardian Nata Atmaja}
\affiliation{Research Center for Quantum Physics, National Research and Innovation Agency (BRIN)\\
 Kompleks PUSPIPTEK Serpong, Tangerang 15310, Indonesia.}
 \emailAdd{muly031@brin.go.id, emir002@brin.go.id, ardi002@brin.go.id}

\abstract{
    We consider a most general $SU(2)$ Yang-Mills-Higgs model consist of terms up to quadratic in first-derivative of the fields, that is the generalized $SU(2)$ Yang-Mills-Higgs with additional scalars-dependent coupling $\theta$-term. Using the BPS Lagrangian method we try to find Bogomolnyi's equations for BPS monopoles and dyons by taking most general BPS Lagrangian density. We obtain more general Bogomolnyi's equations and a relation between all scalars dependent couplings.  From these equations we can see there is a family of BPS monopole solutions parameterized by a real constant $\gamma$, while for BPS dyons there is an additional parameter which is the coupling of $\theta$-term. Interestingly even for a single BPS dyon we find the value of $\theta$-term's coupling only gives additional contribution to electric charge of BPS Dyons, which is in accordance with Witten's result in Phys.Lett.B 86 (1979), and thus can determine whether we get BPS monopoles or BPS dyons. 
}

\begin{document}
\maketitle

\section{Introduction}
Dyons, which are a natural extension of monopoles, are essentially monopoles that also carry a nonzero electric charge. Schwinger~\cite{ Schwinger:1969ib} initially proposed them as an alternative to quarks,  and also their quantum mechanical properties were first explored by Zwanziger \cite{Zwanziger:1969by, Zwanziger:1968rs}.  Similar to monopoles, dyons naturally arise in non-Abelian gauge theory. The first demonstration of monopole existence was provided by Polyakov \cite{Polyakov1974} and ’t Hooft \cite{Hooft1974} in the $SU(2)$ Yang–Mills–Higgs (YMH) model. Later, Julia and Zee \cite{Julia:1975ff} showed that dyons can also exist within the same framework. 

Prasad and Sommerfield~\cite{Prasad:1975kr} then proposed explicit solutions for the 't Hooft-Polyakov monopoles and Julia-Zee dyons by considering a special limit of the model. These solutions satisfy a set of first-order differential equations, known as Bogomolny's equations, derived by Bogomolny~\cite{Bogomolny:1975de}. The solutions to these sets of equations satisfy the nontrivial static energy bound, which is proportional to the topological charge. 

Recent studies of monopoles and dyons have introduced new features and dynamics. Some of these investigations have focused on modifying the SU(2) YMH model. One approach involves introducing additional degrees of freedom along with additional global symmetries \cite{Shifman2015}. Another study proposed modifying the SU(2) YMH model by incorporating scalar field-dependent coupling into each kinetic term \cite{Casana:2012un}. This modified version, referred to as the generalized SU(2) YMH model, allows monopoles to possess internal structures \cite{Bazeia2018}. Recently, in \cite{Atmaja:2018cod}, Atmaja proposed a method to generalize BPS monopoles and dyons in the SU(2) YMH model by introducing non-boundary terms. These terms can subsequently be determined based on the constraints of the system. It is later found that BPS dyons exist in the generalized SU(2) YMH model where the couplings depend explicitly on the Higgs field, \(\Phi\) \cite{Atmaja:2020iti}. 

The couplings of the generalized model, mentioned above, depend explicitly on \(\text{Tr}\left[\Phi^2\right]\), which implies that the generalized model still preserves CP symmetry. In this type of model, it is known that both the electric and magnetic charges of dyons are integers, due to the Dirac quantization \cite{dirac1931quantised,Schwinger:1969ib,Julia:1975ff}. However, this property is modified when CP symmetry is broken. One of the earliest proposal of non-Abelian models which exhibit CP violation is the topological Yang-Mills theory where the Lagrangian contains additional CP-violating term, also known as the \(\theta\)-term \cite{witten1979dyons,witten1988topological}. This CP-violating term is proportional to \(\epsilon^{\alpha\beta\mu\nu}F_{\alpha\beta}F_{\mu\nu}\), where \(\epsilon^{\alpha\beta\mu\nu}\) is the Levi-Civita tensor and \(F_{\mu\nu}\) is the component of field-strength two-form. An interesting property of this model is that, although the contribution of the CP-violating term vanishes in the energy-momentum tensor due to its topological nature, the dyon charge is modified with correction that is proportional to the coupling constant of the CP-violating term \cite{witten1979dyons}. This allows the charge to be non-integer. Thus, it is interesting to study the properties of the dyons within the generalized version of this CP-violating theory. 

In this work, we study the BPS dyons of the generalized SU(2) YMH model with a CP-violating term whose coupling constant is given by \(H\). The BPS equations are calculated using the BPS Lagrangian method, similar to the method used in \cite{Atmaja:2020iti}. The corresponding BPS Lagrangian is modified to accommodate the CP-violating term.

\section{Generalized $SU(2)$ Yang-Mills-Higgs Model}

We consider the generalized $SU(2)$ Yang-Mills-Higgs model~\cite{Casana:2012un,Atmaja:2018cod,Atmaja:2020iti}\footnote{Here we follow the notations in~\cite{Atmaja:2018cod}.} with additional CP-violating term, whose Lagrangian is given by 
 \begin{equation}\label{gen YMH}
  \mathcal{L}=-{w(|\Phi|)\over 2}\tr\left(F_{\mu\nu}F^{\mu\nu}\right)+\frac{H\left(|\Phi|\right)}{4}\epsilon^{\mu\nu\rho\sigma}\tr\left(F_{\rho\sigma}F_{\mu\nu}\right)+ G(|\Phi|)\tr\left(D_\mu\Phi D^\mu\Phi\right)-V(|\Phi|),
 \end{equation}
where $w,G>0,~V\geq 0,$ and $H$ are functions of scalar fields, with $|\Phi|=2\tr\left(\Phi^2\right)$; $F_{\mu\nu}=\partial_\mu A_\nu-\partial_\nu A_\mu-ie \left[A_\mu,A_\nu\right]$; $D_\mu\equiv\partial_\mu-ie\left[A_\mu,\right]$; and $\mu,\nu=0,1,2,3$ are spacetime indices with metric signature $(+---)$. In terms of components, the gauge and scalar fields are
\begin{equation}
 A_\mu={1\over 2}\tau^a A^a_\mu,\qquad \Phi={1\over 2}\tau^a \Phi^a,
\end{equation}
with $a=1,2,3$ and $\tau^a$ are the Pauli matrices. Let us define $F_{0i}\equiv E_i={1\over2}\tau^a E^a_i$ and ${1\over 2}\epsilon_{ijk}F_{jk}\equiv B_i={1\over2}\tau^a B^a_i$, we may then rewrite the Lagrangian density \eqref{gen YMH} to be
\begin{eqnarray}
\mathcal{L}&=&w~ \tr\left(E_i^2-B_i^2\right)+2H~\tr\left(E_iB_i\right)+G~\tr\left(D_0\Phi^2-D_i\Phi^2\right)-V.
\end{eqnarray}

\section{General BPS Lagrangian}
Let us consider a most general BPS Lagrangian as suggested in~\cite{Atmaja:2020iti} as follows
\begin{eqnarray}\label{Lbps gen 2}
 \mathcal{L}_{BPS}&=&X_0+X_1 \tr\left(E_iD_i\Phi\right)+X_2 \tr\left(B_iD_i\Phi\right)+X_3 \tr\left(E_iB_i\right)\nonumber\\
 &&+X_4 \tr\left(D_i\Phi\right)^2+X_5 \tr\left(D_0\Phi\right)^2+X_6 \tr\left(E_i\right)^2+X_7 \tr\left(B_i\right)^2,
\end{eqnarray}
where $X_0,\ldots,X_7$ are auxiliary functions of $|\Phi|$. With this general BPS Lagrangian,
\begin{eqnarray}\label{L-Lbps exps}
\mathcal{L}-\mathcal{L}_{BPS}&=&BE_i+BB_i+BD_0\Phi+BD_i\Phi-V-X_0
\end{eqnarray}
where
\begin{subequations}
\begin{eqnarray}
 BE_i&=&\left(w-X_6\right) \tr\left(E_i-\frac{ \left(X_3-2H\right) B_i+ X_1 D_i\Phi}{2 (w-X_6)}\right)^2,\label{BPSE gen 2}\\
 BB_i&=&-\frac{4 (w-X_6) (w+X_7)+\left(X_3-2H\right)^2}{4 (w-X_6)}\tr\left(B_i+\frac{2 X_2 (w-X_6)+X_1 \left(X_3-2H\right)}{4 (w-X_6) (w+X_7)+\left(X_3-2H\right)^2}D_i\Phi \right)^2,\label{BPSB gen 2}\nonumber\\
 \\
 B\Phi_t&=&\left(G-X_5\right)\tr\left(D_0\Phi\right)^2,\label{BPSD0P}\\
 B\Phi_i&=&\left(\frac{(2 w X_2+X_1 \left(X_3-2H\right)-2 X_2 X_6)^2}{4 (w-X_6) \left(4 (w-X_6) (w+X_7)+\left(X_3-2H\right)^2\right)}-G-X_4-\frac{X_1^2}{4 (w-X_6)}\right)\tr\left(D_i\Phi\right)^2.\label{BPSDiP}\nonumber\\
\end{eqnarray}
\end{subequations}
Here we require $X_6\neq w, X_7\neq {(X_3-2H)^2\over 4(X_6-w)}-w,$ and $X_5\neq G$ for Bogomolnyi's equations to exist. In the BPS limit, $\mathcal{L}-\mathcal{L}_{BPS}=0$, we may extract Bogomolnyi's equations from \eqref{BPSE gen 2}, \eqref{BPSB gen 2}, and \eqref{BPSD0P}, which are given by
\begin{subequations}\label{Bogomolnyi-1}
   \begin{eqnarray}
 E_i&=&{2X_1\left(w+X_7\right)-X_2 \left(X_3-2H\right) \over 4\left(w-X_6\right)\left(w+X_7\right)+\left(X_3-2H\right)^2}D_i\Phi \equiv\alpha D_i\Phi,\label{BE-E gen}\\
 B_i&=&-{2X_2\left(w-X_6\right)+X_1 \left(X_3-2H\right) \over 4\left(w-X_6\right)\left(w+X_7\right)+\left(X_3-2H\right)^2}D_i\Phi \equiv\beta D_i\Phi\label{BE-B gen},\\
 D_0\Phi&=&0.\label{BE-D0Phi}
\end{eqnarray} 
\end{subequations}
We can rewrite $X_1$ and $X_2$, in terms of $\alpha$ and $\beta$, as follows
\begin{equation}
 X_1= 2\left(w-X_6\right)\alpha-\left(X_3-2H\right)\beta,\qquad\qquad X_2=-\left(X_3-2H\right)\alpha-2\left(w+X_7\right)\beta.
\end{equation}
From the Bogomolny's equations (\ref{BE-E gen}) and (\ref{BE-B gen}), it is easy to see that $D_i\Phi\neq 0$, otherwise those Bogomolny's equations will be trivial, and thus no Bogomolnyi's equation extracted from \eqref{BPSDiP}, but instead we must set
\begin{equation}\label{X4}
 X_4=-G-\frac{X_1^2}{4 (w-X_6)}+\frac{(2 w X_2+X_1 \left(X_3-2H\right)-2 X_2 X_6)^2}{4 (w-X_6) \left(4 (w-X_6) (w+X_7)+\left(X_3-2H\right)^2\right)}
\end{equation}
which, in terms of $\alpha$ and $\beta$, can be simplified to
\begin{equation}\label{X4s}
 X_4=-G-\left(w-X_6\right)\alpha^2 +\left(w+X_7\right)\beta^2+\left(X_3-2H\right)\alpha\beta.
\end{equation}
The remaining terms in the right hand side of equation (\ref{L-Lbps exps}) is zero such that $X_0=-V$.

\subsection{Constraint Equations}
In finding solutions to the Bogomolnyi's equations we must also consider constraint equations which are Euler-Lagrange equations of the BPS Lagrangian density \eqref{Lbps gen 2}.
\subsubsection{The Euler-Lagrange equations}
The Euler-Lagrange equations of the BPS Lagrangian (\ref{Lbps gen 2}):
for $\Phi$,
\begin{eqnarray}
&-&2{\partial X_0\over\partial|\Phi|}\Phi+2{\partial X_1\over\partial|\Phi|}\left[\tr\left(\Phi\partial_i\Phi\right)E_i-\tr\left(E_iD_i\Phi\right)\Phi\right]+{X_1\over 2} D_iE_i\nonumber\\
&+&2{\partial X_2\over\partial|\Phi|}\left[\tr\left(\Phi\partial_i\Phi\right)B_i-\tr\left(D_i\Phi B_i\right)\Phi\right]+\frac{X_2}{2}D_iB_i-2{\partial X_3\over\partial|\Phi|}\tr\left(E_i B_i\right)\Phi\nonumber\\
&-&2{\partial X_4\over\partial|\Phi|}\left[\tr\left(D_i\Phi\right)^2\Phi-2\tr\left(\Phi\partial_i\Phi\right)D_i\Phi\right]+X_4 D_iD_i\Phi\nonumber\\
&-&2{\partial X_5\over\partial|\Phi|}\left[\tr\left(D_0\Phi\right)^2\Phi-2\tr\left(\Phi\partial_0\Phi\right)D_0\Phi\right]+X_5 D_0D_0\Phi\nonumber\\
&-&2{\partial X_6\over\partial|\Phi|}\tr\left(E_i\right)^2\Phi-2{\partial X_7\over\partial|\Phi|}\tr\left(B_i\right)^2\Phi=0,
\end{eqnarray}
for $A_i$,
\begin{eqnarray}
 &&2{\partial X_1\over\partial|\Phi|}\tr\left(\Phi\partial_0\Phi\right)D_i\Phi+{X_1\over 2}\left(D_0D_i\Phi-ie\left[E_i,\Phi\right]\right)\nonumber\\
 &-&2{\partial X_2\over\partial|\Phi|}\epsilon_{ijk}\tr\left(\Phi\partial_j\Phi\right)D_k\Phi-{X_2\over 2}\left(\epsilon_{ijk}D_jD_k\Phi-ie[\Phi,B_i]\right)\nonumber\\
 &+&{X_3\over 2}\left(D_0B_i-\epsilon_{ijk}D_jE_k\right)\nonumber\\
 &+&2{\partial X_3\over\partial|\Phi|}\left[\tr\left(\Phi\partial_0\Phi\right)B_i-\epsilon_{ijk}\tr\left(\Phi\partial_j\Phi\right)E_k\right]+ieX_4\left[\Phi,D_i\Phi\right]\nonumber\\
 &+&4{\partial X_6\over\partial|\Phi|}\tr\left(\Phi\partial_0\Phi\right)E_i+X_6 D_0E_i\nonumber\\
 &-&4{\partial X_7\over\partial|\Phi|}\epsilon_{ijk}\tr\left(\Phi\partial_j\Phi\right)B_k-X_7\epsilon_{ijk}D_jB_k=0,
\end{eqnarray}
for $A_0$,
\begin{eqnarray}
 &-&2{\partial X_1\over\partial|\Phi|}\tr\left(\Phi\partial_i\Phi\right)D_i\Phi-{X_1\over 2} D_iD_i\Phi\nonumber\\
 &-&2{\partial X_3\over\partial|\Phi|}\tr\left(\Phi\partial_i\Phi\right)B_i-{X_3\over 2}D_iB_i+ieX_5\left[\Phi,D_0\Phi\right]\nonumber\\
 &-&4{\partial X_6\over\partial|\Phi|}\tr\left(\Phi\partial_i\Phi\right)E_i-X_6 D_iE_i=0.
\end{eqnarray}

\subsubsection{The Bianchi identity}
The equations of motion for the gauge fields are not only given by the Euler–Lagrange equations, but also by the Bianchi identity,
\begin{eqnarray}
    \epsilon^{\sigma\rho\mu\nu}D_\rho F_{\mu\nu}=0~,
\end{eqnarray}
which can be devided into two equations: \\
\begin{eqnarray}
    D_i B_i&=&0~,\label{DiBi}\\
     2D_0 B_i&=&\epsilon_{ijk}D_{[j}E_{k]}~.\label{D0Bi}
\end{eqnarray}
In the BPS limit, using the Bogomolny’s equations \eqref{Bogomolnyi-1}, the equation \eqref{DiBi} can be written as
\begin{eqnarray}\label{Bianchi-DDphi}
    D_iD_i\Phi=-\frac{4\beta'}{\beta}\tr\left(\Phi D_i\Phi\right)D_i\Phi~,
\end{eqnarray}
while the equation \eqref{D0Bi}, for static cases, becomes
\begin{eqnarray}
    \label{Bianchi 2}
\alpha'\epsilon_{ijk}\tr\left(\Phi D_{[j}\Phi\right)D_{k]}\Phi=0~.
\end{eqnarray}

\subsection{Energy Momentum Tensor}
The Lagrangian density (\ref{gen YMH}) has the following stress-energy-momentum tensor:
\begin{equation}
 T_{\mu\nu}=2 G~\tr\left(D_\mu\Phi D_\nu\Phi\right)-2 w~ \tr\left(F_{\lambda\mu}{F^\lambda}_\nu\right)-\eta_{\mu\nu}\mathcal{L}.
\end{equation}
One can easily show that the momentum components are zero in the BPS limit. On the other hand the stress density tensor components in the BPS limit is
\begin{equation}
    T_{ij}=2\left(G-w(\alpha^2+\beta^2)\right)\tr\left(D_i\Phi D_j\Phi\right)-\delta_{ij}\left(G-w(\alpha^2+\beta^2)\right)\tr\left(D_k\Phi\right)^2.
\end{equation}
As argued in~\cite{Atmaja:2020iti}, the stable BPS monopoles and dyons are related to vanishing stress density tensor, and hence $G=w(\alpha^2+\beta^2)$. From now on we will only consider these stable BPS monopoles and dyons.
 
\subsection{Bogomolnyi's Equations}
Substituting the Bogomolny's equations \eqref{Bogomolnyi-1}, $X_4$ solution \eqref{X4s}, and the Bianchi identities (\eqref{Bianchi-DDphi} and\eqref{Bianchi 2}) into the constraint equations for $A_i$, we obtain
\begin{equation}
\alpha'\left(G-w \left(\alpha ^2+\beta ^2\right)\right) \left[D_i\Phi,\Phi\right]=0,\label{CE gen Ai}
\end{equation}
while the constraint equations for $A_0$ are now
\begin{eqnarray}\label{CE gen A0}
    \alpha\left(\left(\alpha w\right)'-\alpha w \frac{\beta'}{\beta}+H'\beta\right)\tr\left(\Phi D_i\Phi\right)D_i\Phi=0~.
\end{eqnarray}
From these two equations we can conclude that either \(\alpha=0\) or, for $\alpha\neq 0$,
\begin{eqnarray}
   \label{Gw} G-w \left(\alpha ^2+\beta ^2\right)&=&0,
    \\ \label{aw}\left(\alpha w\right)'-\alpha w \frac{\beta'}{\beta}+H'\beta&=&0.
\end{eqnarray}
Since we consider only stable solutions, thus equation \eqref{Gw} is trivial. The remaining constraint equations, for $\Phi$, are simplified to
\begin{eqnarray}\label{CE gen Phi}
  &-&X_0'\Phi+\left(G'-w'\left(\alpha^2-\beta^2\right)-2H'\alpha\beta\right) \tr\left(D_i\Phi\right)^2\Phi\notag\\
  &-&2\left(G'-G\frac{\beta'}{\beta}\right) \tr\left(\Phi D_i\Phi\right)D_i\Phi=0,
\end{eqnarray}
Since $\Phi,D_i\Phi\neq0$ this constraint equation can be seen as polynomial equation in powers of $\Phi$ and $D_i\Phi$. So we can solve this equation by setting all their "coefficients" to zero. From the first term we get $X_0=$ constant. It is suggested to take $X_0=0$ which implies $V\to0$. This is inline with the BPS limit condition for BPS monopoles and dyons in \cite{Prasad:1975kr} . For the second and the third term of \eqref{CE gen Phi}, we can substitute \eqref{Gw} and \eqref{aw} into \eqref{CE gen Phi} to show that both "coefficients" are equivalent. Hence, we can take 
\begin{eqnarray}\label{con1}
   G'-G\frac{\beta'}{\beta}=0
\end{eqnarray}
whose solution is
\begin{eqnarray}\label{beta}
    \beta=C_\beta G,
\end{eqnarray}
with $C_\beta$ is an integration constant.
We can rearrange equation \eqref{aw} in the form
\begin{eqnarray}
    \frac{(\alpha w)'}{\alpha w}-\frac{\beta'}{\beta}+\frac{H'\beta}{\alpha w}=0~,
\end{eqnarray}
whose solution is given by
\begin{eqnarray}\label{alpha}
   \alpha=\frac{(C_H-2H)\beta}{2w}~.
\end{eqnarray}
substituting the results in \eqref{beta} and \eqref{alpha} into \eqref{Gw}, we get a relation between the couplings,
\begin{eqnarray}\label{G}
    G=\frac{1}{wC_\beta^2\left(1+\frac{(C_H-2H)^2}{4w^2}\right)}~.
\end{eqnarray}
Comparing with the result in~\cite{Atmaja:2020iti}, the constants ($C_\beta$ and $C_H$) can be fixed by taking $C_\beta=\cos(\gamma)$ and $C_H=2\tan(\gamma)$, with $\gamma$ is an arbitrary constant, as such
\begin{eqnarray}\label{G_f}
    G=\frac{w}{w^2\cos^2(\gamma)+(\sin(\gamma)-H\cos(\gamma))^2}~.
\end{eqnarray}
In this way we can express $\beta$ in terms of $H$ and $w$ as follows
\begin{eqnarray}\label{beta-Hw}
    \beta=\frac{w\cos(\gamma)}{w^2\cos^2(\gamma)+(\sin(\gamma)-H\cos(\gamma))^2},
\end{eqnarray}
while for $\alpha$ is written as
\begin{eqnarray}\label{alpha-Hw}
   \alpha=\frac{\sin(\gamma)-H\cos(\gamma)}{w^2\cos^2(\gamma)+(\sin(\gamma)-H\cos(\gamma))^2}~.
\end{eqnarray}

\section{Generalized BPS Monopoles and Dyons}
We consider a simple case where $H=H_0$ with $H_0$ is a real constant.
\subsection{BPS monopoles: $\alpha=0$}
First, let us consider the case when $\alpha=0$ or $E_i=0$, which corresponds to the BPS monopole scenario. In this case, from Eq. \eqref{alpha}, the BPS monopole solution can be obtained if $H_0=\tan(\gamma)$  Consequently, the Bogomolny's equations become
\begin{equation}
E_i=0,\qquad B_i=\cos(\gamma)G~D_i\Phi,\qquad D_0\Phi=0, 
\end{equation}
where the relation between the scalar dependent couplings is now $Gw={1\over\cos^2(\gamma)}$. These equations are more general compared to the result in~\cite{Atmaja:2018ddi}. Here, unlike in~\cite{Atmaja:2018ddi}, the constant $\gamma$ is still not fixed to $\gamma=0$ or $\pi$. So there is a family of BPS monopole solutions, for fixed $G$, parameterized by a real constant $\gamma$. The presence of $\theta$-term modifies the Bogomolny's equations and determines whether the solutions are BPS monopoles, with $H_0=\tan(\gamma)$, or are BPS dyons, with $H_0\neq\tan(\gamma)$.

\subsection{BPS Dyons: $\alpha\ne 0$}
\begin{figure}[h]
    \centering
    \includegraphics[width=0.5\linewidth]{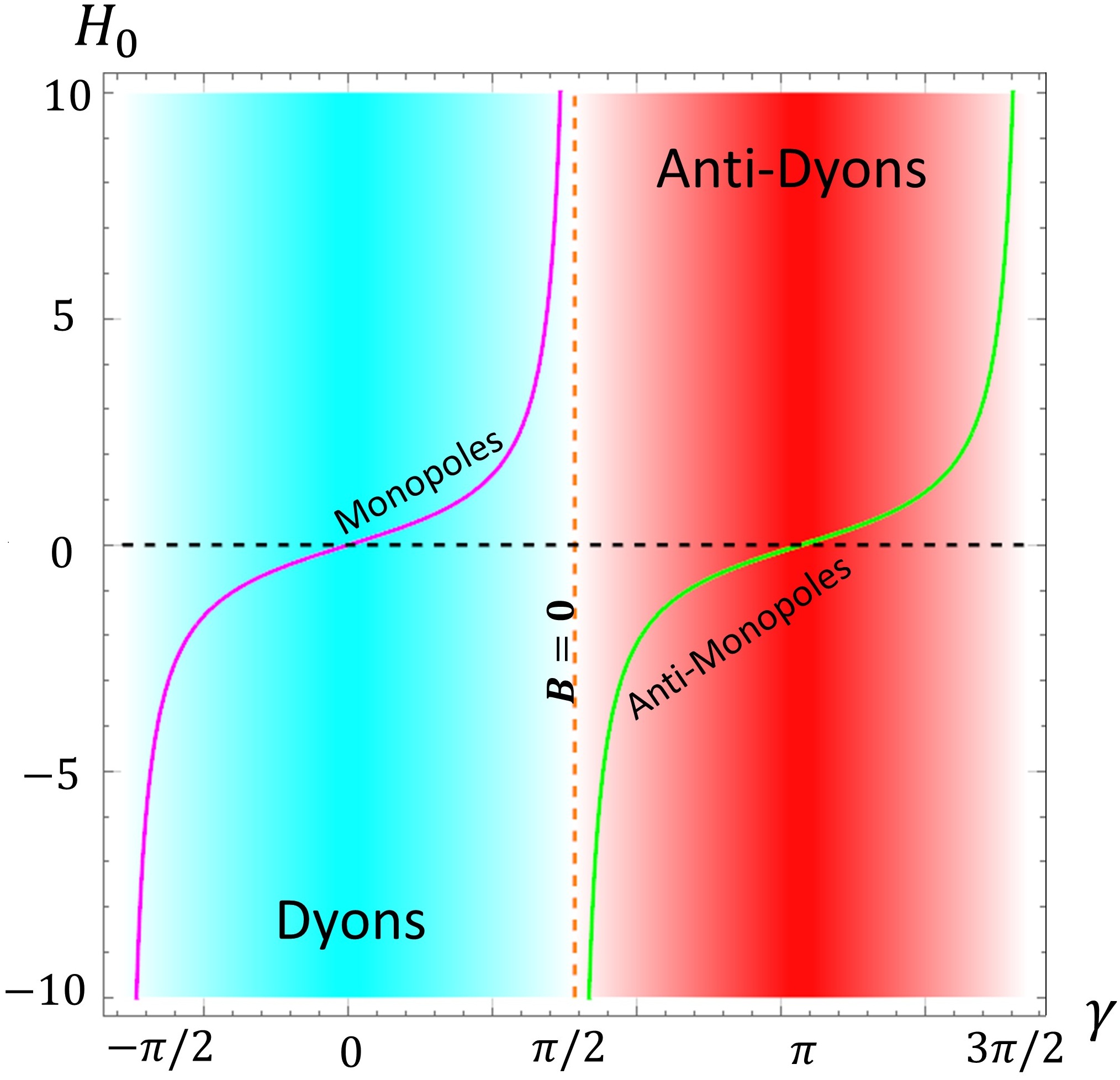}
    \caption{Parameter space of the CP-violation coupling, \(H_0\). We have Dyon solutions for the cyan region and anti-Dyon solutions for the red region. The solid curves represent the special values of \(H_0\) for monopole (magenta) and anti-monopole (green) solutions. The dashed orange line is the limit where \(B_i=0\).}
    \label{fig:Par-Space-CP-Violate}
\end{figure}
In this case a constant $H_0\neq\tan(\gamma)$ and thus Bogomolny's equations are given by
\begin{equation}\label{Final BE}
E_i=\left(\sin(\gamma)-H_0\cos(\gamma)\right)\frac{G}{w}~D_i\Phi,\qquad B_i=\cos(\gamma)G~D_i\Phi,\qquad D_0\Phi=0,\qquad V=0, 
\end{equation}
with
\begin{eqnarray}
    G=\frac{w}{w^2\cos^2(\gamma)+(\sin(\gamma)-H_0\cos(\gamma))^2}~.
\end{eqnarray}
If $H_0=0$ then we get back the Bogomolny's equations and relation between the couplings as in~\cite{Atmaja:2018ddi}. In Fig. \ref{fig:Par-Space-CP-Violate} we plot the parameter space \((H_0,\gamma)\) to show whether a set of values of \(H_0\) and \(\gamma\) produce a dyon or a monopole. According to the BPS equation \eqref{Final BE}, we can see that the transformation form a dyon to the anti-dyon is related the transformation \(\gamma\rightarrow\gamma+\pi\). This can be also  be observed in Fig. \ref{fig:Par-Space-CP-Violate}, since shifting the \(\gamma\) of a dyon solution in the cyan region by \(\pi\) gives us its anti-dyon in the red region. This property can also be found in its monopole limit, \(E_i=0\), represented by the magenta curve. Shifting this curve with respect to \(\gamma\) by \(\pi\) gives us the anti-monopole curve, which has exactly the same profile, but located inside the red region. Since the transformation \(\gamma\rightarrow\gamma+\pi\) inverts the sign of the magnetic field, this transformation implies that the sign of the topological charge is also inverted.

\begin{figure}
   \centering
    \includegraphics[width=0.8\linewidth]{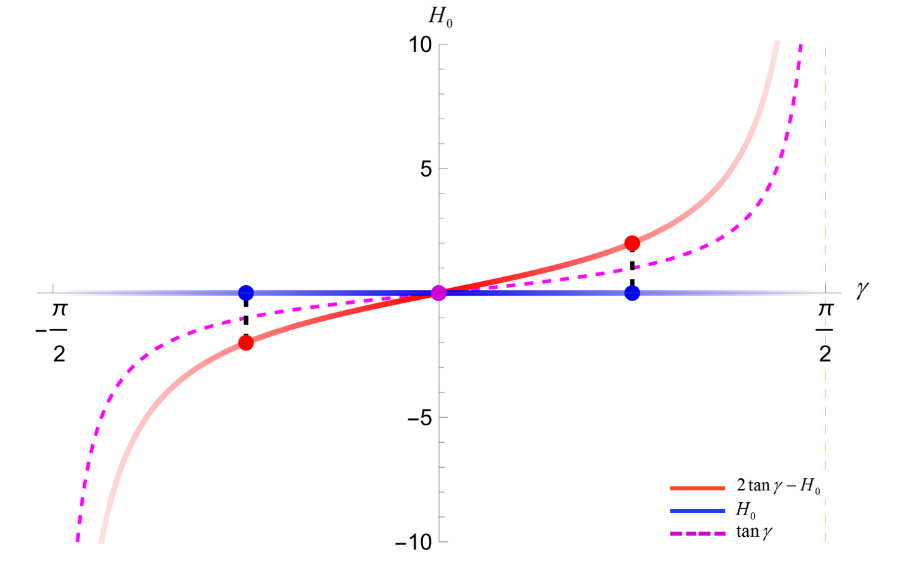}
    \caption{The blue line denotes BPS dyons for $H_0=0$ while the red lines represents BPS dyons with the same magnetic charge, but opposite electric charge. The magenta dashed curve corresponds to BPS monopoles. The figure also illustrates  some fixed \(\gamma\)'s, highlighting the location of BPS dyons, represented by blue dots, and their corresponding BPS dyons, represented by red dots. The relative distance between these dots, represented by black dashed line, describes how much we must shift $H_0$ to change from BPS dyons to their corresponding BPS dyons.}
   \label{fig:enter-label}
\end{figure}
\begin{figure}
   \centering
    \includegraphics[width=0.8\linewidth]{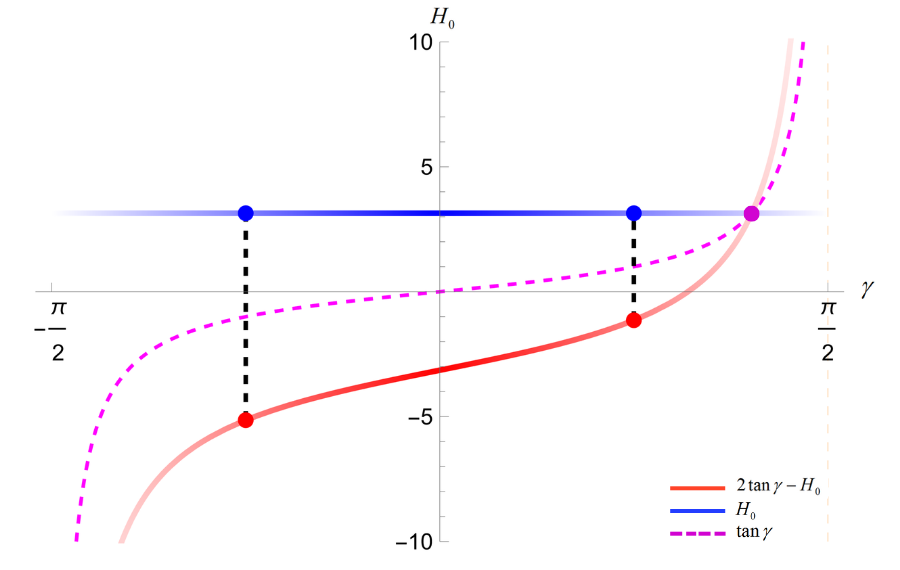}
    \caption{The blue line denotes BPS dyons for $H_0=\pi$ while the red lines represents BPS dyons with the same magnetic charge, but opposite electric charge. The magenta dashed curve corresponds to BPS monopoles. The figure also illustrates  some fixed \(\gamma\)'s, highlighting the location of BPS dyons, represented by blue dots, and their corresponding BPS dyons, represented by red dots. The relative distance between these dots, represented by black dashed line, describes how much we must shift $H_0$ to change from BPS dyons to their corresponding BPS dyons.
}
   \label{fig:enter-label}
\end{figure}
\begin{figure}
   \centering
    \includegraphics[width=0.8\linewidth]{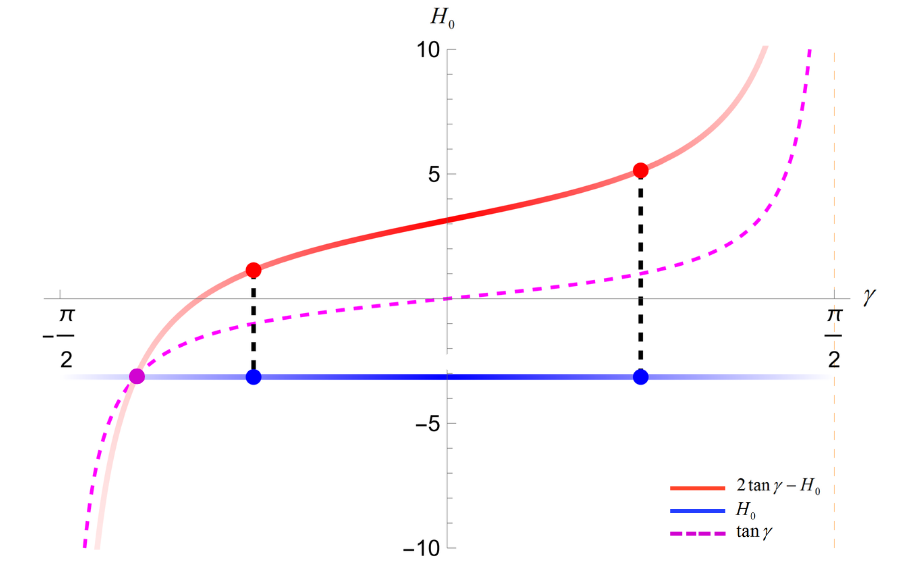}
    \caption{The blue line denotes BPS dyons for $H_0=-\pi$ while the red lines represents BPS dyons with the same magnetic charge, but opposite electric charge. The magenta dashed curve corresponds to BPS monopoles. The figure also illustrates  some fixed \(\gamma\)'s, highlighting the location of BPS dyons, represented by blue dots, and their corresponding BPS dyons, represented by red dots. The relative distance between these dots, represented by black dashed line, describes how much we must shift $H_0$ to change from BPS dyons to their corresponding BPS dyons.}
   \label{fig:enter-label}
\end{figure}

\section{Conclusions and Outlooks}
We have shown using the most general BPS Lagrangian density \eqref{Lbps gen 2} results in similar Bogomolnyi's equations as obtained in \cite{Atmaja:2020iti}, by scaling $\alpha\to\alpha/w$ and $\beta\to\beta/w$. However our result is slightly different. Here surprisingly we still have undetermined auxiliary fields $X_3, X_5, X_6,$ and $X_7$. Although it seems we can take any solution to these auxiliary fields, there are some restrictions to these solutions in order for the Bogomolnyi's equations to be regular such as $X_5\neq G, X_6\neq w,$ and $ X_7\neq {(X_3-2H)^2\over 4(X_6-w)}-w$. One should notice that the results in~\cite{Atmaja:2020iti} correspond to taking $X_3=X_5=X_6=X_7=0$ that do not violate those restrictions. Nevertheless, we do not need to know explicit solutions to these auxiliary fields since they are canceled out and do not show up explicitly at the end of computation.

We obtain similar Bogomolnyi's equations \eqref{Final BE} for the magnetic fields $B_i$, in terms of $G$, as in~\cite{Atmaja:2020iti}. This would imply the total energy of BPS dyon solutions of the Bogomolnyi's equations \eqref{Final BE} is proportional to the topological charge as shown in~\cite{Atmaja:2020iti}. However Bogomolnyi equations \eqref{Final BE} for the electric fields $E_i$ get a different multiplication factor compared to the results in~\cite{Atmaja:2020iti}. There is additional term in the multiplication factor  due to the presence of $\theta$-term, $\left(\sin(\gamma)-H_0\cos(\gamma)\right)$. This would imply electric charge of the BPS dyons get additional contribution from the $\theta$-term coupling $H_0$, which is in accordance to the result found by Witten~\cite{witten1979dyons}. The values of $H_0$, besides $\gamma$, also determine whether the solutions are BPS monopoles or dyons. For particular values of $H_0=\tan(\gamma)$, there exist only BPS monopole solutions, while any other values will give us BPS dyons.  

Without $\theta$-term, by taking \(H_0=0\), the only way to get BPS dyons with opposite  electric field,  \(E\rightarrow-E\) while keeping \(B\rightarrow B\), is to tune the parameter \(\gamma\to-\gamma\). However such transformation can only be done at the field equation level and is the implication of \(CP\) symmetry. Another way to do it is by turning on the $\theta$-term with a particular value $H_0=2\tan(\gamma)$. In general tuning the $\theta$-term coupling, 
\begin{equation}
    H_0\rightarrow2\tan(\gamma)-H_0,
\end{equation}
will result in changing electric charge of the BPS dyons to its opposite value, \((E,B)\to(-E,B)\).

The case where $H$ is not constant, or depends on spatial coordinates, is quite interesting. The electric fields are given by
\begin{equation}
    E_i(\vec{r})=\left(\sin(\gamma)-H(\vec{r})\cos(\gamma)\right)\frac{G}{w}~D_i\Phi.
\end{equation}
The electric fields of BPS dyons will vary on space. There is a possible configuration where values of the electric fields are divided into three regions (positive, zero, negative). Although the electric charge of BPS dyons depends on contribution of the electric fields in the whole space, this configuration of the electric fields is different from the normal configuration where the electric fields are positive/negative in the whole space. In this configuration another BPS dyon may be trapped inside a finite region where the values of electric fields are zero. We limit our discussion in this article for the case of constant $H$. Nevertheless, should we took $H$ to be constant in the beginning, we would not be able to see the effect of $\theta$-term in the electric charge of BPS dyons since the constraint equations \eqref{CE gen A0} and \eqref{CE gen Phi} depend on derivative of $H$ only.

\section*{Acknowledgements}
M.M and E.S.F is supported by the BRIN 2024 Postdoctoral Research Program.


\medskip
\bibliographystyle{utphys}
\bibliography{testBiB}

\end{document}